\documentclass[aps,prb,twocolumn,notitlepage,showpacs,amsmath,amstex,amssymb,citeautoscript,longbibliography,floatfix,a4paper,twocolumn]{revtex4-1}
\pdfoutput=1
\usepackage[english]{babel}
\usepackage{letltxmacro}
\usepackage{latexsym}
\LetLtxMacro{\ORIGselectlanguage}{\selectlanguage}
\makeatletter
\DeclareRobustCommand{\selectlanguage}[1]{%
  \@ifundefined{alias@\string#1}
    {\ORIGselectlanguage{#1}}
    {\begingroup\edef\x{\endgroup
       \noexpand\ORIGselectlanguage{\@nameuse{alias@#1}}}\x}%
}
\newcommand{\definelanguagealias}[2]{%
  \@namedef{alias@#1}{#2}%
}
\makeatother
\definelanguagealias{en}{english}
\definelanguagealias{English}{english}
\usepackage{graphicx}
\usepackage{amsmath}
\usepackage{amsfonts}
\usepackage{amssymb,bbding}
\usepackage{bm}
\usepackage{dsfont}
\usepackage{float}
\usepackage{braket}
\usepackage{hyperref}
\hypersetup{
    bookmarks=false,         % show bookmarks bar?
    unicode=false,          % non-Latin characters in AcrobatÕs bookmarks
    pdftoolbar=false,        % show AcrobatÕs toolbar?
    pdfmenubar=true,        % show AcrobatÕs menu?
    pdffitwindow=false,     % window fit to page when opened
    pdfstartview={FitH},    % fits the width of the page to the window
    pdftitle={Duality approach to clustering in quantum 3-XORSAT models},    % title
    pdfauthor={R. Medina, M. Serbyn},     % author
    pdfsubject={},   % subject of the document
    pdfcreator={},   % creator of the document
    pdfproducer={}, % producer of the document
    pdfkeywords={}, % list of keywords
    pdfnewwindow=true,      % links in new window
    colorlinks=true,       % false: boxed links; true: colored links
    linkcolor=black,          % color of internal links (change box color with linkbordercolor)
    citecolor=blue,        % color of links to bibliography
    filecolor=magenta,      % color of file links
    urlcolor=blue           % color of external links
}

\newcommand{\be}{\begin{equation}}
\newcommand{\ee}{\end{equation}}
\newcommand{\bea}{\begin{eqnarray}}
\newcommand{\eea}{\end{eqnarray}}

\renewcommand{\H}{A}% matrix in XORSAT problem
\newcommand{\Hcl}{{H_\text{c}}}
\newcommand{\Hq}{H_\text{q}}
\newcommand{\nO}{q}
\newcommand{\nG}{g}% number of generations/levels in the tree graph
\begin{document}
\title{Duality approach to quantum annealing of the 3-XORSAT problem} 
\author{Raimel Medina}
\affiliation{IST Austria, Am Campus 1, 3400 Klosterneuburg, Austria}
\author{Maksym Serbyn}
\affiliation{IST Austria, Am Campus 1, 3400 Klosterneuburg, Austria}
\date{\today}
\begin{abstract}
Classical models with complex energy landscapes represent a perspective avenue for the near-term application of quantum simulators. Until now, many theoretical works studied the performance of quantum algorithms for models with a unique ground state. However, when the classical problem is in a so-called clustering phase, the ground state manifold is highly degenerate. As an example, we consider a 3-XORSAT model defined on  simple hypergraphs. The degeneracy of classical ground state manifold translates into the emergence of an extensive number of $Z_2$ symmetries, which remain intact even in the presence of a quantum transverse magnetic field. We establish a general duality approach that restricts the quantum problem to a given sector of conserved $Z_2$ charges and use it to study how the outcome of the quantum adiabatic algorithm depends on the hypergraph geometry. We show that the tree hypergraph which corresponds to a classically solvable instance of the 3-XORSAT problem features a constant gap, whereas the closed hypergraph encounters a second-order phase transition with a gap vanishing as a power-law in the problem size. The duality developed in this work provides a practical tool for studies of quantum models with classically degenerate energy manifold and reveals potential connections between glasses and gauge theories. 
\end{abstract}
\maketitle
\section{Introduction}
The quantum adiabatic algorithm~\cite{Farhi_2000}, which can be viewed as a generalization of quantum annealing~\cite{qsoptimization, QA_1994_Finnila, QA_1998_Nishimori, QA_1999_Brooke, QA_and_QAC_2006}, was considered as a perspective quantum algorithm since early days of quantum computing. In this algorithm, the solution of a classically hard combinatorial optimization problem~\cite{Mezard} is mapped onto a problem of finding a ground state of a classical spin Hamiltonian. Such ground state is in turn obtained by initializing a quantum spin system in a ground state of a simple quantum Hamiltonian and then adiabatically interpolating between the quantum and classical Hamiltonians. The success of this algorithm, which is quantified by the overlap between the final state after the evolution and the ground state, is guaranteed, provided the spectrum features a finite gap throughout the adiabatic evolution, see Refs.~\onlinecite{QAC_rop_2013, QA_review_2015,AQC_review_2018, QA_review_2020} for recent reviews. 

The performance of the algorithm was studied theoretically for several optimization problems~\cite{Zamponi2010,Young&Zamponi2012}. Remarkably, in many cases the gap was shown to vanish polynomially or even exponentially in the problem size~\cite{Zamponi2010,Young&Zamponi2012}, giving evidence of the phase transition encountered in the annealing process. The majority of models studied to date featured a \emph{unique} ground state. While such problems are convenient for numerical studies, in many interesting combinatorial problems one often encounters a degenerate space of solutions. Classical problems with many possible solutions, where some are similar to each other, while others are globally different, are said to be in a ``clustering phase''~\cite{Mezard_2002}. Classical optimization problems in the clustering phase correspond to the spin Hamiltonians with \emph{degenerate} ground state manifold, a situation that is often explicitly ruled out in studies of quantum adiabatic algorithm performance. 

In this work we specifically focus on classical optimization problem with degenerate space of solutions. To this end, we use the ``exclusive-or" satisfiability (XORSAT) problem~\citep{xorsat,xorsat_2} for studies of quantum algorithm performance in clustering phase.   XORSAT is equivalent to a boolean linear algebra problem, hence it is easily verifiable and solvable in satisfiable cases. Restricting to the case where  each exclusive or condition involves exactly 3 variables, we obtain so-called 3-XORSAT problem, that maps onto a classical spin Hamiltonian with three-spin interactions specified by a certain hypergraph. This spin model was studied in the literature, where the existence of clustering phase was established for random hypergraphs ensembles~\cite{Mezard_2002, xorsat_2}. 

We focus on particular instances of the 3-XORSAT problem, which provide an example of classically solvable instances, yet feature a large degeneracy in the space of solutions. We show that such degeneracy in the solution space can be recast into the emergence of a set of $Z_2$ conserved charges that persists in the quantum model. To restrict the problem to a particular sector, we generalize the duality introduced in Ref.~\onlinecite{Young&Zamponi2012}. The application of duality to spin model on a tree hypergraph results in an Ising-type model, facilitating numerical and analytical understanding. In particular, we establish that the 3-XORSAT model on a tree hypergraph does not feature a phase transition, guaranteeing the success of the quantum adiabatic algorithm. On the other hand, the closure of the tree hypergraph leads to an emergence of the second-order phase transition encountered over the course of adiabatic evolution.

The structure of this paper is organized as follows. In Sec.~\ref{sec:2} we briefly review the 3-XORSAT problem as well as the quantum adiabatic algorithm. In Sec.~\ref{sec:3} we illustrate the duality mapping using specific instances of the 3-XORSAT problem. For each of these instances, we find the dual Hamiltonian, as well as discuss its energy spectrum and minimal gap dependence with system size. We conclude in Sec.~\ref{sec:4} with a brief discussion of our results and a summary of interesting directions for future work. 

\begin{figure*}[t]
\begin{center}
\includegraphics[width=1.99\columnwidth]{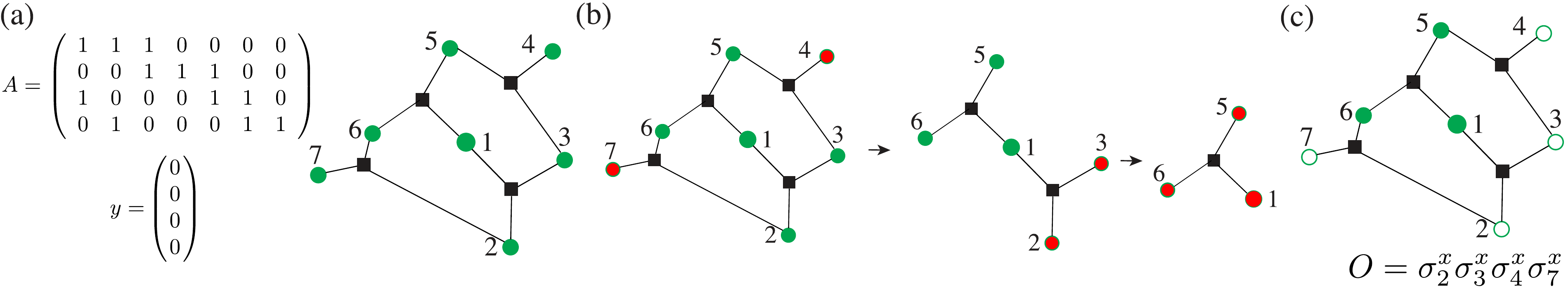}\\
\caption{ \label{Fig:intro}
(a) Matrix $\H$ that specifies 3-XORSAT problem with $N = 7$ variables and $M = 5$ conditions, and corresponding hypergraph where vertices shown by green dots denote spins and black squares are the edges that correspond to three-spin interaction terms. 
(b) Illustration of leaf removal algorithm that can find the solution of the classical problem. Starting from the original hypergraph in (a) at each step one removes spins that enter in just one interaction (equivalently, are included only in one edge). At the first step one removes spins $4$ and $7$.  Then we can remove either spins $2, 3$ or spins $5, 6$.  At the last step all three remaining spins can be removed. (c)  Simultaneous flip of spins $2,3,4,7$ (white filled circles) does not change the energy of the system. Such degeneracy corresponds to the operator $O$ that commutes with the classical Hamiltonian. 
}
\end{center}
\end{figure*}

\section{Classical and quantum 3-XORSAT model}{\label{sec:2}}
In this section, we introduce the classical 3-XORSAT model and associated spin Hamiltonian. We briefly review the application of the so-called ``leaf removal algorithm''~\cite{Mezard_2002} to find the solution of a classical problem and highlight the emergent degeneracy of the classical energy landscape. Finally, we discuss the application of the quantum adiabatic algorithm for finding the ground state of the classical 3-XORSAT model. We show that even though the degeneracy of the classical energy landscape is lifted in the presence of a transverse field, a set of commuting integrals of motion remains.

\subsection{Classical 3-XORSAT\label{Sec:XORSAT-class}}

Classical 3-XORSAT problem~\cite{xorsat} consists in finding the arrangements of binary variables $x_1,\ldots x_N$ that satisfy the set of $M$ distinct ``exclusive-or'' (XOR) clauses with only three variables participating in each condition. Using equivalence between XOR operator and binary addition, we can rewrite the XOR clause  $x_1\oplus x_2\oplus x_3 = b$ where $b=0,1$ as $x_1+x_2+x_3  = b\, {\rm mod}\, 2$. This allows to map a 3-XORSAT problem onto a system of linear equations: 
\be
\label{eq:XORSAT}
\H \cdot x = y \quad \text{(mod 2)},
\ee
where $\H$ is a $M \times N$ matrix and $y$ is a $M-$component vector with binary entries, $\H_{ai} \in \{0,1\}$, $y_a \in \{0,1\}$. Since we restricted to clauses with only three variables, each row of the matrix $\H$ contains exactly three ones with all other entries being zero, see example in Fig.~\ref{Fig:intro}(a). Determining whether the Boolean system of equations~\eqref{eq:XORSAT} admits an assignment of the Boolean variables satisfying all the equations constitutes the decision version of the 3-XORSAT problem. In general, one is also interested in the set of solutions and its size. Throughout this work our focus will be on quantum annealing approach to finding the solution of the XORSAT problem.

The 3-XORSAT problem defined by means of a linear system of equations with $N$ variables and $M$ equations can be naturally mapped to the problem of energy minimization for ensemble of $N$ classical spins, $\sigma^z_i$, with $M$ three-spin interactions~\cite{Mezard}. Defining $\sigma^z_i = (-1)^{x_i}$ and $J_a = (-1)^{y_a}$ one can demonstrate that solution of Eq.~(\ref{eq:XORSAT}) corresponds to a zero-energy ground state of the following classical Hamiltonian:
\begin{equation}\label{eq:XORHam}
\Hcl= \sum_{\alpha=1}^{M} (1-J_\alpha \sigma^z_{i_\alpha}\sigma^z_{j_\alpha}\sigma^z_{k_\alpha}).
\end{equation}
In case when the system of equations~(\ref{eq:XORSAT}) does not admit a solution that satisfies all conditions (it is said to be UNSAT), the ground state of the $\Hcl$ corresponds to a bit assignment that violates the minimal possible number of conditions.

The 3-XORSAT problem and corresponding classical Hamiltonian are fully fixed by the pair of $(\H,y)$, or, equivalently the choice of three-spin interactions and a value of couplings, $J_\alpha = \pm 1$.  Interactions between spins can be conveniently visualized using the hypergraph, where vertices correspond to spins, and edges (which now join three spins, hence these are in fact hyperedges) correspond to interactions. A particular instance of the 3-XORSAT problem and corresponding hypergraph is illustrated in Fig.~\ref{Fig:intro}(a).

The hypergraph representation provides a visual way to find the solution to the 3-XORSAT problem. The so-called leaf removal algorithm~\cite{Leaf_removal} is illustrated in Fig~\ref{Fig:intro}(b) and consists of removing the spins that enter only in a single interaction. The insight is that if a given spin, say $\sigma^z_7$, appears in the Hamiltonian only once, e.g. in the term $\sigma^z_2\sigma^z_6\sigma^z_7$ for the chosen example, we can always satisfy the corresponding interaction term by adjusting the value of $\sigma^z_7$. Thus we are allowed to erase this spin and the corresponding interaction term. Iterating such search and removal of spins that enter a single interaction term (so-called leaves) on the hypergraph is the essence of the leaf removal procedure. This procedure halts if one removes all vertices and edges as is shown in Fig.~\ref{Fig:intro}(b). This case corresponds to an instance of the 3-XORSAT problem that is completely solvable by the leaf removal algorithm. Another alternative is when in the process of iterating leaf removal procedure one fails to find any leaves. The leaf removal algorithm halts at such an instance and the remaining hypergraph is typically dubbed a ``glassy core''~\cite{Mezard_2002}, see an example of such hypergraph in Fig.~\ref{Fig:closedL}(a).

During the iterative process of the leaf removal algorithm, one may encounter instances when more than two spins participating in a given interaction term are simultaneously removed, see Fig~\ref{Fig:intro}(b). When such interaction edge and two associated spins are removed a degeneracy emerges. In the example in Fig.~\ref{Fig:intro}(b) we remove simultaneously $\sigma^z_2$ and $\sigma^z_3$, hence flipping these spins simultaneously does not affect the energy of the given interaction edge. At the level of the full Hamiltonian, such instances lead to an emergence of global degeneracies: in the example that we show the total energy does not change if one flips spins 2, 3, 4, and 7. Depending on the geometry of the problem, one may encounter many such degeneracies with their number being a finite fraction of the total number of spins --- this is characteristic of the so-called clustering phase ~\cite{Mezard_2002,xorsat_2}. Some of this degeneracy though originates from the structure of the glassy core, which typically does not have a unique solution (UNSAT) but instead has multiple degenerate ground states.

\subsection{Solving 3-XORSAT with quantum adiabatic algorithm\label{Sec:QAA}}

One approach to finding the ground state of the classical Hamiltonian~(\ref{eq:XORHam}) or, equivalently, to finding the bit assignment that violates the smallest possible number of equations in~(\ref{eq:XORSAT}) is provided by quantum adiabatic algorithm~\cite{Farhi_2000}. Supposing that classical Hamiltonian~(\ref{eq:XORHam}) can be implemented on a quantum simulator, we initialize the system in the ground state of a quantum paramagnet Hamiltonian 
\begin{equation}\label{Eq:Hq}
\Hq = -\sum_{i=1}^N\sigma^x_i,
\end{equation}
 and evolve this state under the following time-dependent Hamiltonian:
\be
\label{eq:paramH}
H_T(t) = (1-\frac{t}{T})\Hq + \frac{t}{T}\Hcl,
\ee
from time $t=0$ to $T$. According to the adiabatic theorem, if $T$ is sufficiently large and $\Hq$ and $\Hcl$ do not commute with each other, the quantum simulator will remain with high fidelity in the ground state for all times, resulting in a preparation of the ground state of $\Hcl$ at time $T$.

The running time $T$, depends on the energy spectrum of $H_T(t)$. In particular, the time required for preparing the ground state with high fidelity is bounded from below by the inverse square of the minimum gap encountered throughout the time evolution, $T \gg {\max_t |V_{10}(t)|}/{[\min_t \Delta(t)]^2}$. Here the gap is defined as a difference between the energy of the ground state and the first excited state, $\Delta(t) = E_1-E_0$, and $V_{10} = \langle 0| \partial_t H(t)|1\rangle$ is the matrix element of the time-dependent part of the Hamiltonian between ground state $\ket{0}$ and first excited state $\ket{1}$. Due to this bound, many theoretical studies of the efficiency of quantum adiabatic algorithm focus on the behavior of the minimum gap of $H_T(t)$~\cite{van_Dam_2001,QAA_failures_Farhi_2005}. 

\subsection{Behavior of gap and degeneracies}

The behavior of the gap for the so-called 3-regular 3-XORSAT Hamiltonian, where each spin enters in exactly three interaction terms, was considered previously~\cite{Zamponi2010,Young&Zamponi2012}. It was found that the system goes through a first-order quantum phase transition, displaying an exponential decrease of the gap with system size. However, these studies were restricted to the instances of the classical 3-XORSAT problem that do not have any degeneracy in the ground state. These instances are said to have \emph{unique satisfying assignment}, and their consideration simplifies the study of the gap behavior~\cite{Zamponi2010,Young&Zamponi2012}. For the 3-XORSAT problem defined on a 3-regular ensemble of random hypergraphs in the $N\to \infty$ these instances form a non-zero fraction ($\sim$ 0.285) of the set of all instances~\cite{Zamponi2010}. Yet, the behavior of instances that have degenerate ground state manifold was not studied.

In this work we (to the best of our knowledge) provide the first results relative to systems with degenerate ground states. We consider instances where degeneracy of the ground state originates from the existence of simultaneous spin flips that do not change the energy of the classical Hamiltonian (see discussion in Section~\ref{Sec:XORSAT-class}). We note, that the ground state may have additional degeneracy due to the problem being UNSAT, which is not considered here. If simultaneous flipping of spins $\sigma^z_{i_1}\to-\sigma^z_{i_1}$,\ldots, $\sigma^z_{i_k}\to-\sigma^z_{i_k}$  does not change the energy of the system, the following operator
\begin{equation}\label{Eq:O-generic}
O = \sigma^x_{i_1}\sigma^x_{i_2}\ldots \sigma^x_{i_k},
\end{equation}
commutes with classical Hamiltonian,  $[O,\Hcl] = 0$. Since the quantum Hamiltonian, $\Hq$, contains only $\sigma^x$ terms, any such operator also commutes with the full $H_T(t)$, 
$$[O, H_T(t)] = 0,$$ 
for any $t$, thus corresponding to an Abelian $Z_2$ \emph{symmetry} present in the system. Moreover, as we mentioned above, many typical instances of the 3-XORSAT problem may contain a possibly extensive number of distinct operators $\{O_l\}$, $l=1,\ldots \nO$ that commute not only with the Hamiltonian but also among themselves. 

The presence of $q$ distinct Abelian symmetries leads to spectral degeneracy only for the classical Hamiltonian, i.e.\ only for $H_T(t)$ at $t=T$. However, although these symmetries do not give rise to spectral degeneracy when $t<T$, their presence fragments the $2^N$-dimensional Hilbert space of model~(\ref{eq:paramH}) into $2^q$ distinct sectors, each labeled by $\pm 1$ eigenvalues of corresponding $O_l$ operator. The full Hamiltonian assumes block-diagonal form when written in the basis that diagonalizes operators $\{O_l\}$, 
\be\label{eq:blockdecomp}
H_T(t) = \bigoplus_{\alpha=1}^{2^q} {H}_\alpha(t),
\ee
where $\alpha$ runs over all $2^q$ blocks. 

The unitary evolution preserves the symmetries of the Hamiltonian. This implies that the search for the minimum gap is performed inside the block ${H}_{\alpha}(t)$, which contains the initial state, $|\psi(0)\rangle$. Due to the reduced dimensionality of $H_\alpha$, we can perform exact numerical calculations for a wide range of system sizes.

One of the main results of this work is the \emph{duality transformation} which allows to explicitly obtain the form of the Hamiltonian ${H}_{\alpha}(t)$ restricted to a given sector.  In the next section, we introduce this duality transformation using specific examples. This duality allows us to readily study the behavior of the gap even in presence of extensive degeneracies in the system and understand the fate of quantum adiabatic algorithm. 

\begin{figure*}[t]
	\begin{center}
		\includegraphics[width=1.95\columnwidth]{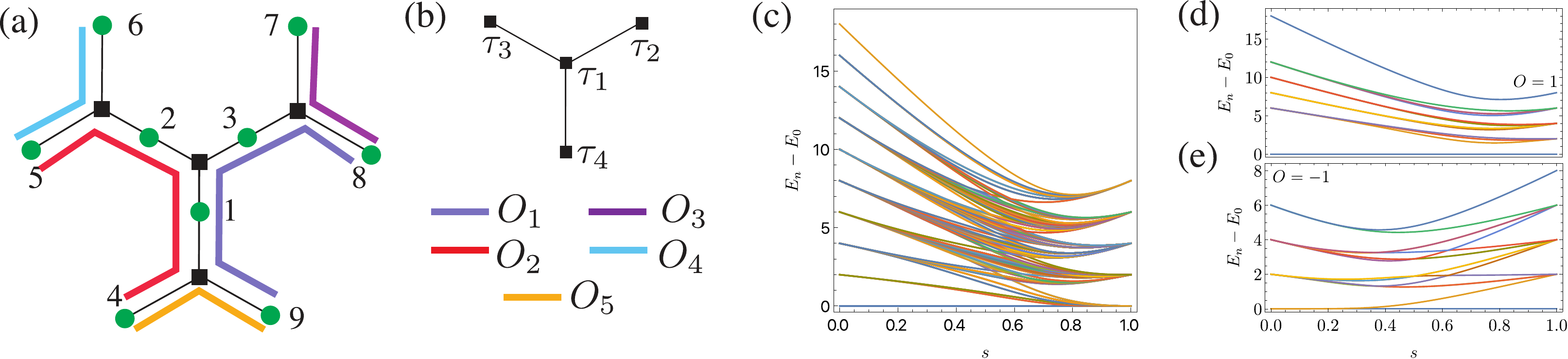}
		\caption{ \label{Fig:hushimi}
			(a) Example of the Hushimi tree at the level $\nG=2$. A convenient choice of the set of independent conserved quantities is shown by colored lines with different lines corresponding to individual conserved quantities, for instance $O_1 = \sigma^x_1\sigma^x_8\sigma^x_9$. (b) Dual degrees of freedom live on the tree hypergraph with $\nG-1$ generations. (c) Evolution of low lying spectrum as a function of parameter $s=t/T$ reveals many crossings and large degeneracy in the spectrum of classical Hamiltonian at $s=1$. (d)~Spectrum of dual Hamiltonian in the sector where all charges $O_l=1$ has only avoided crossings demonstrating that application of duality resolves all symmetries. (e) Spectrum of the dual Hamiltonian in the sector $O_l=-1$, where the dual model has an additional emergent $Z_2$ symmetry, that is manifested in degeneracy of ground state manifold of the dual model for small values of~$s$. }
	\end{center}
\end{figure*}

\section{Duality approach to quantum 3-XORSAT model}{\label{sec:3}}

As discussed above, the duality provides a natural approach to quantum 3-XORSAT Hamiltonian in presence of conserved quantities. In this section we illustrate duality using specific instances of 3-XORSAT model, whereas in the Appendix~\ref{app:general_duality} we formulate the duality using the language of linear algebra which allows to apply such transformation to the 3-XORSAT problem on arbitrary graphs in an efficient manner. 

\subsection{Duality for tree hypergraph \label{sec:OpenLattice}}
The structure of degeneracies in the 3-XORSAT model is determined by its connectivity. While often the 3-XORSAT model is considered on random graphs~\cite{Zamponi2010, Young&Zamponi2012}, below we consider an instance of the 3-XORSAT problem that is fully solvable by the leaf removal algorithm. In particular, we consider a \emph{tree hypergraph} that may be thought of as a toy example of the structure of the leaves of the generic 3-XORSAT instances. We find that the dual Hamiltonian is an Ising model and obtain that the energy gap remains constant in the thermodynamic limit. 
\subsubsection{Degeneracies and conserved charges}

We consider the 3-XORSAT problem on the tree hypergraphs with connectivity 2 and a varying number of generations. An example of tree hypergraph shown in Fig~\ref{Fig:hushimi}(a) has $g = 2$ generations of spins and contains $N = 3(2^g-1) = 9$ vertices and $M = 4$ edges. Any such tree hypergraph corresponds to a trivial solvable instance of 3-XORSAT: application of leaf removal algorithm completely removes all vertices and results in a solution.

In the process of a leaf removal iteration, one always encounters pairs of spins that belong to the same edge and are removed simultaneously. As explained in Sec.~\ref{Sec:XORSAT-class}, this leads to degeneracies. The tree hypergraph with $g$ generations is characterized by $\nO = 3\cdot2^{g-1}-1$ independent $Z_2$ charges. For the particular hypergraph in Fig.~\ref{Fig:hushimi}(a) this formula yields $\nO = 5$ charges, which are shown by different colors in Fig.~\ref{Fig:hushimi}. A given symmetry sector can be fixed by specifying the eigenvalues of all these charges. In particular, the ground state of the quantum part of the annealing Hamiltonian, $\Hq$ in Eq.~(\ref{Eq:Hq}), $\ket{\psi(0)} = \ket{\rightarrow\ldots \rightarrow}$ corresponds to the values of all charges $O_l = 1$. We are interested in performing a duality transformation that restricts the Hamiltonian to a particular symmetry sector. Taking into account that the ratio between the number of independent charges and the number of spins $q/N$ tends to the value of $1/2$ in the thermodynamic limit $g,N\to\infty$, the duality is capable of drastically reducing the Hilbert space dimension from $2^N$ to approximately~$2^{N/2}$.

\subsubsection{Dual Hamiltonian}
We explicitly construct the duality, by defining spins $\tau$ that live at the edges of the hypergraph, see Fig.~\ref{Fig:hushimi}(b). The $\tau^x$ operators are expressed via original spins as: 
\begin{equation}\label{Eq:tri-x}
\tau_{(ijk)}^x = \sigma_i^z\sigma_j^z\sigma_k^z,
\end{equation}
where $\tau_{(ijk)}^x$ is the dual spin located at the edge that was connecting spins $(i,j,k)$. In order to simplify notations, we label the edges and dual spin operators $\tau_\alpha$ by greek indices as in Fig.~\ref{Fig:hushimi}(b);  for instance, $\tau^x_{\alpha=1} = \tau^x_{(123)}  = \sigma^z_1\sigma^z_2\sigma^z_3$. This mapping converts the classical Hamiltonian, $\Hcl$ in Eq.~(\ref{eq:XORHam}) into the simple sum of $\tau^x_i$ operators,
\begin{equation}\label{Eq:Hcl-dual}
\tilde \Hcl = -\sum_{\alpha \in V} J_\alpha \tau^x_\alpha,
\end{equation}
where we omitted a constant term from Eq.~(\ref{eq:XORHam}). Tilde emphasizes that this Hamiltonian acts in the Hilbert space of $\tau$-spins and index $\alpha$ runs over all vertices of the dual graph, Fig.~\ref{Fig:hushimi}(b), denoted as $V$. 

Similar to duality applied to discrete Abelian gauge theories~\cite{fradkin_2013}, the relation between the $\tau^z$ and $\sigma^x$ is non-local. The $\tau^z$ operators are defined via product of $\sigma^x$ operators on the path from a certain ``root vertex'',
\begin{equation}\label{Eq:tri-z}
\tau^z_{\alpha} = \prod_{m\in\text{path to }\alpha} \sigma^x_m.
\end{equation}
This root vertex is chosen as $i=9$ in Fig.~\ref{Fig:hushimi}(a). Then for the graph in Fig.~\ref{Fig:hushimi}(b) we have: $\tau^z_{\alpha=4}  = \tau^z_{(149)} = \sigma^x_9$, $\tau^z_{1}  = \sigma^x_9\sigma^x_1$, $\tau^z_{2} =\sigma^x_9\sigma^x_1\sigma^x_3$, and $\tau^z_3 =\sigma^x_9\sigma^x_1\sigma^x_2$. This construction will result in the simple expression for original spins, $\sigma^x_1 = \tau^z_{1}\tau^z_{4}$, unless they are located at the boundary of the graph. Thus, for the bulk spins the dual $\tilde{\Hq}$ of $\Hq$ coincides with an Ising model on a tree

However, the situation is different for the boundary spins. In order to obtain the expression for $\sigma^x_i$ at the boundary, one must use the existence of the conserved charges. For example, the spin $\sigma^x_4$ cannot be expressed via the product of any of the four $\tau^z_\alpha$ operators. However, we observe that $\sigma^x_4 = (\sigma^x_4\sigma^x_9)\sigma^x_9 = O_1\sigma^x_9 = O_1\tau^z_{4}$. Remaining boundary spins $\sigma^x_{i}$ with $i =5,\ldots 8$ can be constructed in a similar way.  Dual spin operators $\tau^{x,z}_\alpha$  defined in such way obey the standard Pauli commutation relations,  $\{ \tau^z_\alpha, \tau^x_\alpha \} = 0$ and $ [\tau^z_\alpha, \tau^x_\beta] = 0$ for $\alpha\neq\beta$. 

Collecting all terms together and denoting $s = t/T$ we obtain the dual of the full Hamiltonian, Eq.~(\ref{eq:paramH}) as: 
\begin{multline}
\label{Eq:H-is-full}
\tilde H_T(s)=
- s\sum_{\alpha \in V} J_\alpha \tau^x_\alpha
-(1-s)\sum_{\langle\alpha \beta\rangle   \in V} \tau_\alpha^z\tau_\beta^z 
\\ 
-(1-s) \sum_{\alpha \in \partial V} h^z_\alpha[O] \tau_\alpha^z.
\end{multline}
The first two lines here correspond to the Ising model on a Cayley tree, see Fig.~\ref{Fig:hushimi}(b). The last line encodes the dependence of duality on the values of conserved charges, and involves only $\tau$-spins at the boundary of the Cayley tree $\partial V$~($\tau^z_{2,3,4}$ in the present example). The effective symmetry-breaking field coupled to boundary spins reads: 
\begin{equation}\label{Eq:hz}
h^z_\alpha[O] =  (1+O_{m_\alpha}) \prod_{m \in \text{path from root}} O_m.
\end{equation}
Here $O_{m_\alpha}$ is the charge that involves only two spins, including $\alpha$, and product is over all charges encountered on the path from the root. For instance, $h_4[O] = 1+O_5$, $h_2[O]=(1+O_3) O_1$ in notation of Fig.~\ref{Fig:hushimi}. 

Remarkably, the first line in the dual Hamiltonian Eq.~(\ref{Eq:H-is-full}) is the Ising model on the Cayley tree with connectivity equal to three. This part of the dual Hamiltonian has global $Z_2$ symmetry $\tau^z\to -\tau^z$ and does not depend on the values of conserved charges (signs of $J_\alpha$ can be removed by the relabeling $\tau^x\to-\tau^x$ in the present case). However, in addition, we also have the second line in Eq.~(\ref{Eq:H-is-full}) that imposes a $Z_2$-symmetry breaking effective field on the dual degrees of freedom at the boundary. The strength of these symmetry-breaking fields depends on the sector of conserved charges as we discuss below.  

\subsubsection{Energy spectrum and minimal gap of dual Hamiltonian}{\label{sec:min_gap_open_lattice}}
To illustrate the advantage of describing the system with the dual Hamiltonian, we show the spectrum of the original Hamiltonian Eq.~(\ref{eq:paramH}) as a function of $s$ in Fig.~\ref{Fig:hushimi}(c). The low-lying energy levels become highly degenerate at $s=1$, corresponding to the degeneracy of the ground state manifold of the classical problem. Moreover, we observe multiple level crossings between eigenstates that belong to different symmetry sectors. The level crossings and degeneracy complicate the determination of the minimal gap encountered throughout the adiabatic algorithm. 

In comparison, Fig.~\ref{Fig:hushimi}(d-e) demonstrates the spectrum of the dual Hamiltonian~(\ref{Eq:H-is-full}) for particular values of conserved charges (also referred to as ``sector'') has much lower complexity. These energy levels are a subset of energy levels shown in Fig.~\ref{Fig:hushimi}(c). The sector of conserved charges is a \emph{property of initial state}. The ground state of the quantum paramagnet $|\rightarrow\rightarrow\ldots \rightarrow\rangle$, is an eigenstate of all $O_m$ operators with eigenvalue $O_m = 1$. Thus, from Eq.~(\ref{Eq:hz}) we obtain a \emph{uniform} magnetic field $h^z_\alpha[O]= 2$ for all $\alpha$ at the boundary of the tree. The presence of this magnetic field leads to a strong breaking of $Z_2$ symmetry that would be otherwise present in the dual Hamiltonian. Hence, it helps to avoid the second-order phase transition in an Ising model, and Fig.~\ref{Fig:hushimi}(d) shows that the finite gap of order one is present for all values of $s$.

\begin{figure}[b]
\begin{center}
\includegraphics[width=0.9\columnwidth]{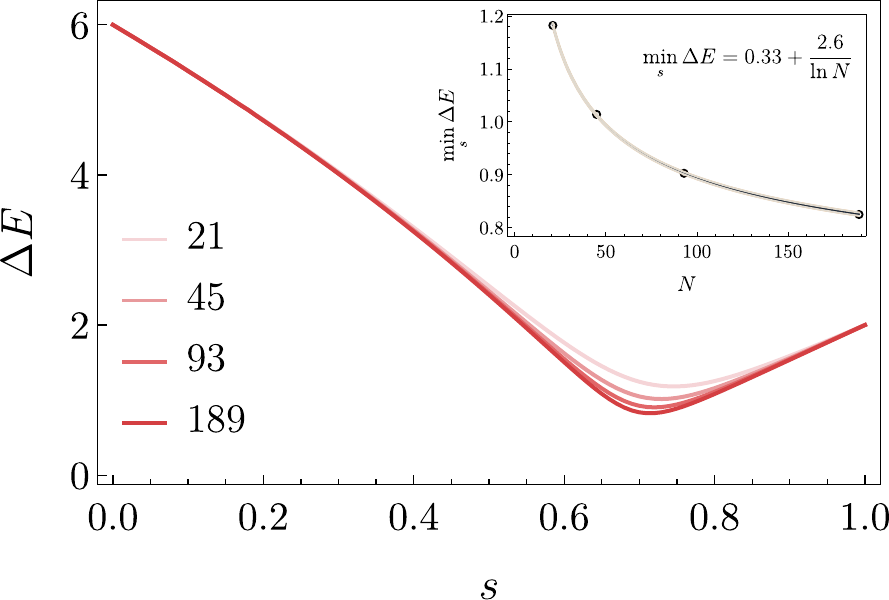}\\
\caption{ \label{Fig:gap-scaling-Ising}
Behavior of energy gap as a function of $s$ for open hypergraphs with different number of generations in the sector $O_l = 1$ demonstrates that gap has minimal value around $s\approx 0.7$. The finite size scaling in the inset shows that the gap approaches constant value in the thermodynamic limit with corrections decaying as $1/\ln N$. Data is obtained with DMRG algorithm implemented in iTensor~\cite{itensor} with bond dimension $\chi=6$ and number of sweeps $n_{\rm{sweeps}}=30$.}
\end{center}
\end{figure}

The duality facilitates the determination of the gap on several levels. First, it decreases the number of degrees of freedom and allows us to study the problem in a smaller Hilbert space. Second, it removes the degeneracies and explicitly resolves all symmetries present in the problem, making the extraction of the energy gap more straightforward. 
As a result, the duality allows us to study the finite-size scaling of the gap for the family of tree hypergraphs with up to $\nG = 6$ generations with $N=189$ spins. We use the density-matrix renormalization-group (DMRG) algorithm to obtain the ground state and energy gap as a function of the parameter $s$. Previous works have studied the transverse field Ising model on the Cayley tree~\cite{MPSCayley,MPSCayley_efficient,Sondhi2008QCavity} with a global symmetry breaking field. In our study, we apply DMRG algorithm to an Ising model with symmetry-breaking fields at the boundary, corresponding to the energy spectrum encountered in the adiabatic algorithm launched from the paramagnetic ground state. The resulting behavior of the gap for different system sizes, $N =3(2^g-1)$ is shown in Fig.~\ref{Fig:gap-scaling-Ising}.

The finite-size scaling of the gap, shown in the inset of Fig.~\ref{Fig:gap-scaling-Ising} reveals that the gap approaches a constant value with corrections that decay logarithmically in the number of spins~$N$. This is consistent with expectations that finite magnetic field applied to all boundary spins (these in the case of the Cayley tree constitute the finite fraction of all spins) destroys the phase transition. The presence of a gap in the thermodynamic limit allows us to conclude that the quantum adiabatic algorithm can efficiently find the ground state of the 3-XORSAT model on the considered hypergraph.

Due to the degeneracy present in this model, one can arrive at the ground state starting from a different initial state which has values of $O_{3,4,5}=-1$ so that the symmetry breaking field vanishes. In the initial spin basis, this correspond to choosing an initial state where the pairs of out-most spins on the boundary triangles have different spin values, i.e., $\sigma_i^x = -1$ for $i=4,6,8$ while $\sigma^x_i=1$ for all remaining value of $i$. In this case, however, we encounter a second-order phase transition as a function of parameter $s$, see Fig.~\ref{Fig:hushimi}(e). This result is in agreement with previous findings~\cite{MPSCayley} of a second-order phase transition at $s_c\approx 0.5733$ which is characterized by a critical correlation length, $\xi = 1/\ln 2$. This peculiar behavior is due to the tree geometry of the lattice, and it is not observed for systems on local lattices, where the correlation length is known to diverge at the critical point.

\subsection{Duality for closure of tree hypergraph}
\begin{figure*}[tb]
		\includegraphics[width=1.99\columnwidth]{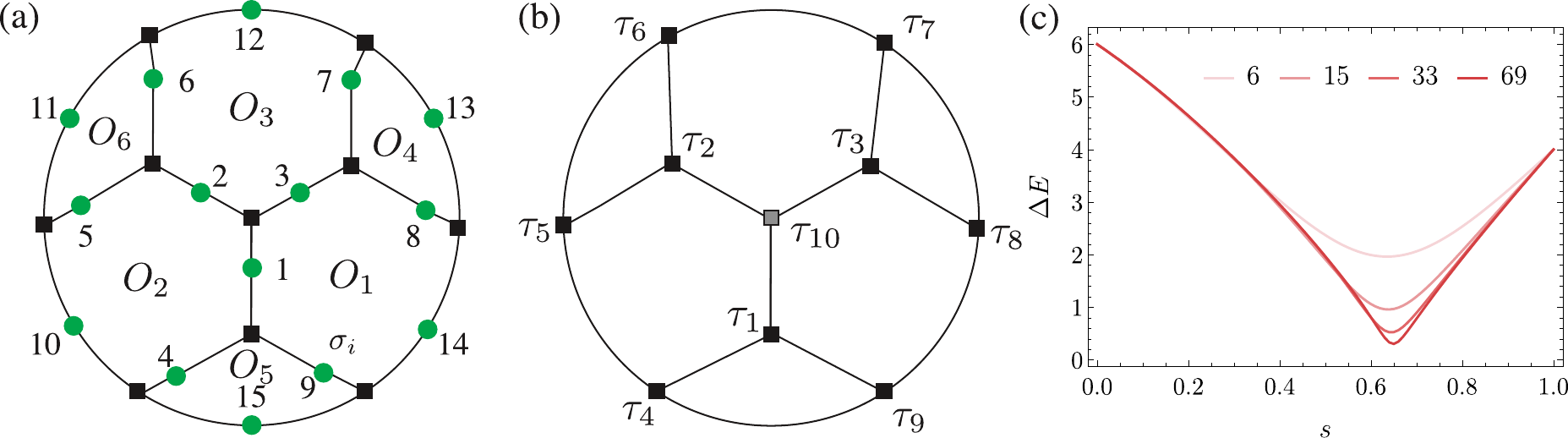}
		\caption{ \label{Fig:closedL}
(a) Closure of the tree hypergraph at level $g=2$ removes the boundary and leads to a 3-XORSAT instance where no spins can be decimated by leaf removal algorithm. The conserved charges labeled by $O_{1,\ldots 6}$ correspond to internal loops of the lattice. (b) Dual degrees of freedom live on the closure of the tree hypergraph. The central $\tau$-spin shown by gray square is redundant. (c) The dependence of the minimal gap on the system size is extracted from DRMG algorithm. }
\end{figure*}
We continue the illustration of the duality by applying it to a hypergraph without a boundary shown in Fig.~\ref{Fig:closedL}(a). This hypergraph can be thought of as the closure of the tree hypergraph considered above. It corresponds to an instance of the 3-XORSAT problem that does not admit a solution by the leaf removal algorithm. Indeed, all spins enter into at least two interaction edges, thus the leaf removal algorithm cannot remove any leaves at all. This second example may be considered as an example of the ``glassy core''~\cite{Mezard_2002}, and the presence of non-trivial loops leads to the appearance of non-local terms in the dual Hamiltonian. Using the duality we will argue that the minimal gap vanishes polynomially in the inverse problem size. 

\subsubsection{Degeneracies and conserved charges}

The closure of the tree hypergraph with $\nG$ generations has $\nO = 3\cdot2^{\nG-1}$ independent conserved quantities. The choice of $O_l$ in Fig.~\ref{Fig:closedL}(a) for the graph with $g=2$ results in six conserved charges that are in one-to-one relation with the spins on the boundary. For example, $O_1= \sigma^x_1\sigma^x_3 \sigma^x_8 \sigma^x_9 \sigma^x_{14}$ includes only one boundary spin $\sigma_{14}$. Given that the total number of spins is $N=3(3\cdot2^{\nG-1}-1)$ in the general case, we expect that the dual Hamiltonian has $N_\tau= 3(2^{\nG}-1)$ spins. For the particular instance of the graph in Fig.~\ref{Fig:closedL}(a) this gives $N=15$ and $N_\tau = 9$.

In comparison with Section~\ref{sec:OpenLattice} here the structure of the ground state manifold is more complicated. In particular, before we ignored the presence of couplings $J_\alpha$ since their value could be always made positive. In the present case, this is not possible anymore. Instead, we find that for any set of the coupling constants $J_\alpha=\pm 1$ it is possible to relabel operators $\sigma^z\to-\sigma^z$, so that either (i) all couplings $J_\alpha = 1$, or (ii) only one coupling is negative, $J_M = -1$, and all remaining couplings are positive. The relabeling procedure does not influence an overall parity, so option (i) is realized if $\prod_{\alpha=1}^{M} J_\alpha = 1$, while (ii) holds when $\prod_{\alpha=1}^{M} J_\alpha = -1$. Below we focus on case (i), where the system has a ground state with energy $E_0 = -M$, where $M$ is the number of interaction edges, or, alternatively, the classical system of equations has an assignment that satisfies all conditions. On the other hand, in case (ii) the system is UNSAT and the ground state energy is $E_0 =M - 2$. Furthermore, for the UNSAT case, the ground state has an additional $M$-fold degeneracy compared to the case (i). We reserve considerations of UNSAT case for future studies.

\subsubsection{Dual Hamiltonian}

To perform the duality transformation, we associate the $\tau$-spins with interaction edges, see Fig.~\ref{Fig:closedL}(b). However, the number of interaction edges is larger than the number of dual spins: this is related to the fact that each $\sigma$ spin enters into 2 interaction edges. Thus, the product over all interaction edges,
$
\prod_{\alpha=1}^{M} \sigma^z_{i_\alpha}\sigma^z_{j_\alpha}\sigma^z_{k_\alpha} = 1,
$
results in an identity operator. We use the same relation Eq.~(\ref{Eq:tri-x}) to define $\tau^x_\alpha$ operator via the product of $\sigma^z_i$ spins in the corresponding interaction edge. The presence of a constraint for the product of all interaction edges allows expressing one of the $\tau$ spins via the remaining operators,
\be \label{eq:constraint}
\prod_{\alpha=1}^{M} \tau^x_\alpha = 1,  
\qquad 
\tau^x_M = \prod_{\alpha =1}^{M-1} \tau^x_\alpha. 
\ee 
While there is a freedom in choosing the `redundant' $\tau$-spin, we fix it to be the central spin, see the shaded square in Fig.~\ref{Fig:closedL}(b). In what follows we do not explicitly express  $\tau_M$ spin via remaining spins to keep notation compact. 

To define $\tau^z_i$ operators we use the central site of the dual lattice as a ``root''. In particular, we define
\be\label{eq:tf-dual-1}
\tau^z_i = \sigma^x_i, \quad \text{for}\quad  i=1,2,3.
\ee
Then, the remaining $\tau^z$ can be written as the product of $\sigma^x_{s \in \mathcal{P}_i}$, where $\mathcal{P}$ correspond to a path in the lattice starting from the site $i=1,2,3$.
In order to write the quantum part of Hamiltonian in the dual basis, we express $\sigma^x_i$ operators via spins $\tau^z_\alpha$. It is straightforward to see that $\sigma^x_i = \tau^z_\alpha\tau^z_\beta$ where edges $\alpha$ and $\beta$ both share the spin $i=1,\ldots, 9$ (basically all spins except the outer layer). For the spins at the outer layer of the graph we again rely on the presence of conserved charges, finding that $\sigma^x_i = O_{l_i}\tau^z_\alpha\tau^z_\beta$ where $O_{l_i}$ is the charge that contains spin $i$. For instance, $\sigma^x_{15} = O_{5}\tau^z_6\tau^z_7$ in notations of Fig.~\ref{Fig:closedL}.

With the above relations we can finally write the expression for the dual Hamiltonian
\begin{multline}\label{eq:DualHam-final}
\tilde H_T(s) = -s\sum^M_{\alpha=1} J_\alpha \tau^x_\alpha
 - (1-s)\sum_{\langle \alpha \beta \rangle} \eta_{\alpha\beta} \tau^z_\alpha \tau^z_\beta\\	
- (1-s) \sum_{\alpha=1}^{3}\tau^z_\alpha,
\end{multline}
where effective couplings between dual spins $\alpha,\beta$ depends on the location of the spin as well as on the value of conserved charges: 
$$
\eta_{\alpha\beta} = \left\{
\begin{array}{@{}l@{\thinspace}l}
	1  & \  \text{if} \ \{ \alpha, \beta \} \notin \partial V,\\
	{O}_{l_{i}} &\   \text{if} \ \{ \alpha, \beta \} \in \partial V, \ \tau^z_\alpha \cap \tau^z_\beta = \sigma^x_i.
\end{array}
\right.
$$

\begin{figure}[b]
\includegraphics[width=0.93\columnwidth]{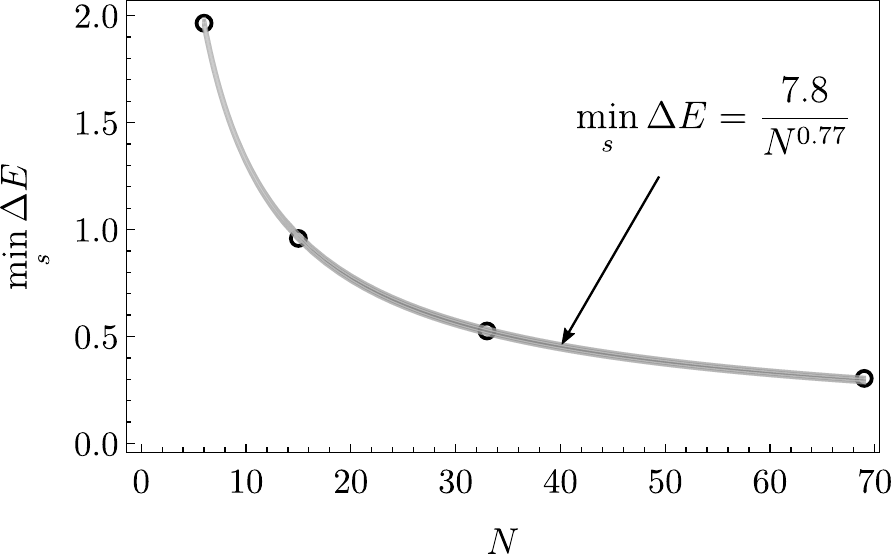}
\caption{ \label{Fig:closedL_mingap_scaling}
The finite size scaling shows that the gap vanishes as a power-law in system size with a coefficient $c=0.77$. Data is obtained with DMRG implemented in iTensor~\cite{itensor} with bond dimension $\chi=8$ and number of sweeps $n_{\rm{ sweeps} } =40$.}
\end{figure}

\subsubsection{Energy spectrum and minimal gap of dual Hamiltonian}
As in the previous case, the value of all conserved charges is fixed by the initial state on the physical basis. The ground state of quantum Hamiltonian leads to all ${O}_l $ having eigenvalue $1$. The dual Hamiltonian in this sector corresponds to Eq.~(\ref{eq:DualHam-final}) with all $\eta_{\alpha\beta}=1$ supplemented by the expression for $\tau^x_M$ via remaining spins, Eq.~(\ref{eq:constraint}). It is interesting to compare Eq.~\eqref{eq:DualHam-final} with Eq.~\eqref{Eq:H-is-full}. One difference is the appearance of a non-local term in Eq.~\eqref{eq:DualHam-final} that is implicitly encoded in $\tau^x_M$ operator. More importantly, in the case of the tree hypergraph, one could obtain a strong symmetry-breaking magnetic field on the boundary by an appropriate choice of conserved charges. This boundary field allowed to eliminate the second-order phase transition, resulting in a finite value of gap even in the thermodynamic limit. In the case of closure of tree hypergraph, the symmetry breaking field is only present for a vanishing fraction of spins (more precisely, three spins in the center for the present gauge choice), resulting in a very different physics as we discuss below. 

In Appendix~\ref{app:ising_closed_lattice} we demonstrate that Eq.~\eqref{eq:DualHam-final} with all $\eta_{\alpha\beta}=1$ is equivalent to the transverse field Ising model on the closed lattice [see Fig.~\ref{Fig:closedL}(b)] in an enlarged Hilbert space that also includes spin $\tau_M$ as a physical degree of freedom,
\be\label{eq:DualHamIsing}
	\tilde H_T(s) = -s\sum_{\alpha=1}^{M} J_\alpha \tau^x_\alpha- (1-s)\sum_{\langle \alpha \beta \rangle} \tau^z_\alpha \tau^z_\beta.
\ee
The behavior of the transverse field Ising model on the closure of the tree hypergraph was not studied before to the best of our knowledge. Due to the presence of loops, the analytical methods applied in the case of the tree hypergraph cannot be used in the present case. Therefore, we resort to numerical simulations, using the same DMRG method as in Sec.~\ref{sec:min_gap_open_lattice}. 

We compute numerically the ground state energy and the gap to the next excited state as a function of $s$, see Fig.~\ref{Fig:closedL}(c). Note, that na\"ively such gap vanishes in the Hamiltonian~(\ref{eq:DualHamIsing}) for values of $s$ close to zero since the model is in symmetry-broken phase. However, as we discuss in Appendix~\ref{app:ising_closed_lattice}, the success of the quantum adiabatic algorithm depends on the gap restricted to the even $Z_2$-symmetry sector. The finite-size scaling of the gap performed for systems with up to $N=69$ spins (corresponding to $M=45$ dual spins) in Fig.~\ref{Fig:closedL_mingap_scaling} shows the gap vanishes as a power-law with system size. This gives strong evidence of a second-order phase transition encountered at $s\approx 0.65$, which can be expected due to the presence of $Z_2$ symmetry in dual Hamiltonian.

\section{Discussion}{\label{sec:4}}

Motivated by the fact that many interesting classical problems have degeneracy in solution space, in this paper we studied the performance of quantum adiabatic algorithms applied to such problems. To this end, we introduced duality as a generic tool that allows us to efficiently target such problems and formulated it using the language of linear algebra in Appendix~\ref{app:general_duality}. In the main text, we demonstrated the application of duality to two different instances of the 3-XORSAT problem.

First, we applied the general duality to the 3-XORSAT problem on a tree hypergraph, which may be thought of as imitating the structure of the leaves of a generic 3-XORSAT instance. Such an instance of the 3-XORSAT problem can be efficiently solved by a classical algorithm in a polynomial time. In Sec.~\ref{sec:OpenLattice} we found that the dual Hamiltonian corresponds to the Ising model with longitudinal magnetic fields at the boundary of the graph. Thus, when starting the annealing process from the paramagnet state the gap saturates to a constant value in the thermodynamic limit with corrections decaying as $1/\ln N$. This implies that the application of the quantum adiabatic algorithm could yield a solution in a finite amount of time, even in the thermodynamic limit. 

As a more general example, we considered a 3-XORSAT problem on the closure of the tree hypergraph, which may be considered as a cartoon picture of a ``glassy core". Despite being non-amenable to the leaf removal algorithm, this instance of the 3-XORSAT problem is still solvable in a polynomial time by a classical algorithm. The presence of non-trivial loops in this geometry translates into the appearance of non-local terms in the dual quantum Hamiltonian. We found that the minimal gap of the annealing Hamiltonian vanishes as a power-law with the problem size, implying the quantum adiabatic algorithm would now require a time that is polynomial in the problem size. 

Despite considering only two toy examples of the 3-XORSAT with extensive degeneracy of classical solution space, the application of duality revealed an interesting connection between the behavior of the minimum gap and the structure of the lattice. In particular, we observed that by closing the boundary of the tree hypergraph the minimum gap changes from being constant in the thermodynamic limit, to decaying as a power law in system size.  This suggests that in the most complex case a first order phase transition may emerge, similarly to other instances of 3-XORSAT with unique ground state considered previously~\cite{Zamponi2010, Young&Zamponi2012}. In addition, duality may be used to obtain useful analytical results for the entanglement spectrum. In particular, we expect the entanglement spectrum of a given subregion to contain information about conserved charges that are supported within the subregion.

More generally, the two considered examples of the tree hypergraph and its closure can be viewed as a basis of perturbation theory, as more typical hypergraphs can be obtained by ``decorating'' the tree hypergraph with additional interactions. In particular, changes to the hypergraph geometry that add additional interaction terms typically break the formerly conserved charges. This would correspond to the introduction of additional non-local degrees of freedom into the dual Hamiltonian. Such an approach can be potentially used to target more complex instances of the 3-XORSAT and possibly relate the problem with classical clustering in the ground state manifold to instances of quantum clustering, that was recently considered in the literature~\cite{Morampudi_PRA_2017}. Additionally, these considerations suggest that frustration that is brought by additional interaction terms naturally corresponds to non-local interactions in the dual language. 

Finally, throughout this work, we focused on the ground state and low-lying excitations of the Hamiltonian used in the quantum adiabatic algorithm to solve the classical 3-XORSAT problem. The study of highly excited states of such Hamiltonians remains an interesting problem, where duality obtained in our work can bring useful insights. In particular, it would be interesting to investigate if these models could allow for a non-ergodic phase similar to the one found in~\cite{Scardicchio_2017_Clustering_QSGlasses}.

\section{Acknowledgments}
We would like to thank S. De Nicola, A. Michaidilis, T. Gulden, Y. Núñez-Fernández, P. Brighi and S. Sack for fruitful discussions and valuable feedback on the manuscript. MS acknowledges useful discussions with E.~Altman, L.~Cugliandolo, and C.~Laumann. We acknowledge support by the European Research Council (ERC) under the European Union's Horizon 2020 research and innovation program (Grant Agreement No.~850899).

\appendix

\section{Generic formulation of duality \label{App:duality}}
\label{app:general_duality}
In the main text, we describe the duality procedure for two particular instances of the 3-XORSAT problem. However, it is desirable to formulate the general procedure of deriving the dual Hamiltonian for general (possibly random) instances of classical 3-XORSAT. In this section, we introduce a general description of the duality transformation that uses the language of linear algebra. 
\subsection{Algorithmic description of duality}
The matrix $\H$ from Eq.~\eqref{eq:XORSAT} is the starting point of our procedure. This formulation, can be seen as an extension of the duality mapping used in~\cite{Young&Zamponi2012} for non-invertible $\H$ matrices. Since $\{\sigma^a_i\}$ and $\{ \tau^b_j \}$ operators, with $a,b\in\{x,y,z\}$ and $i,j\in[1,N]$, belong to different Hilbert spaces, in what follows we will use the symbol $``\equiv"$ to refer to equivalences between then.

\subsubsection{Introducing linear algebra notations}
In contrast to the particular case of duality in~\cite{Young&Zamponi2012}, which required matrix $\H$ to be invertible, here we generally deal with the matrix $\H$ that is not square and thus is not a full-rank matrix. First, let us denote by $r$ the rank (mod 2) of the matrix $\H$, $\mathop{\rm{rank}_2}(\H)=r$. We further define matrices $S_A$ and $S'_A$, which will be used to find $\tau^x$ operators. The matrix $S_A$ contains all linearly independent rows of $\H$,
 \begin{equation}\label{Eq:SA}
S_{\H} = ( v_1 , \ldots, v_r )^T.
\end{equation}
The matrix $S'_A$ contains the remaining rows of $\H$ which by construction can be obtained from those in $S_A$. Hence, this matrix can be written as a linear superposition of the vectors $v_j \in S_A$, 
\begin{equation}\label{Eq:SFA}
S'_A =  F S_A,
\end{equation}
encoded by the $(M-r) \times r$ matrix $F$.

In order to find the $\tau^z$ operators we use a matrix $Z$
\begin{equation}\label{Eq:Z}
Z = (z_i,\ldots, z_r)^T,
\end{equation}
that contains an orthonormal set of vectors $z_i$,  such that $z_j \cdot v_i = \delta_{ij}$. In practice these vectors can be obtained by finding the left-inverse of transposed matrix $S_A$ from Eq.~(\ref{Eq:SA}), $Z \cdot S^T_A = \mathbb{I}_{r \times r}$.

Finally, the conserved charges are associated with the vectors spanning the kernel (mod 2) of $\H$. Since the basis of any linear space is not uniquely defined, we use the following choice of these kernel basis vectors
\be\label{eq:basis_ker}
 \mathcal{O} = ( (S^T_A \cdot Z)_{r+1} + \hat{e}_{r+1}, \ldots (S^T_A \cdot Z)_{N} + \hat{e}_{N})^T, 
\ee
where $\hat{e}_i$ is the unit vector of length $N$ in the $i$-th direction. This choice leads to a particularly simple expression for the dual version of quantum terms $\sigma^x_i$. 

\subsubsection{Finding $\tau^x$ operators}
To construct the $\tau^x_\alpha$ operators we use a set of linearly independent rows of $\H$ matrix contained in matrix $S_\H$, see Eq.~(\ref{Eq:SA}). Each row of $\H$ and $S_\H$ contains exactly three entries that are equal to one since we are dealing with the 3-XORSAT problem. Therefore, we identify
	\begin{equation}\label{Eq:taux-1}
		\tau^x_\alpha \equiv \bigotimes_{l=1}^{N} (\sigma^z_l)^{(S_A)_{\alpha,l}} =\sigma^z_{i_\alpha}\sigma^z_{j_\alpha}\sigma^z_{k_\alpha} , \quad \forall \alpha \in [1, r],
	\end{equation}
where $(i_\alpha, j_\alpha, k_\alpha)$ are indices of non-zero entries of row $\alpha$ of matrix $S_A$.
The remaining $M-r$ rows are then expressed as a linear combination of the vectors in $S_{\H}$ as in Eq.~(\ref{Eq:SFA}). This implies that a product of $\sigma^z$ operators encoded by those vectors can be obtained from $\tau^x$ operators defined above. Specifically, the product of $\sigma^z$'s corresponding to a given row $(S'_A)_l$, where $(S'_A)_l = \sum_{k = 1}^{r} F_{l, k} (S_A)_k$, reads:
	\begin{equation}\label{Eq:taux-2}
        \prod_{k =1}^{r} (\tau^x_k)^{F_{\alpha,k}} \equiv  \sigma^z_{i_\alpha}\sigma^z_{j_\alpha}\sigma^z_{k_\alpha},
	\end{equation}
where we imply that $(\tau^x_k)^{F_{\alpha,k}} = \tau^x_k$ if $F_{\alpha,k}=1$ and $(\tau^x_k)^{F_{\alpha,k}} = 1$ if $F_{\alpha,k}=0$.

Finally, using equations (\ref{Eq:taux-1})-(\ref{Eq:taux-2}) we can express classical Hamiltonian $H_C$ via dual operators as:
\be
\label{eq:sigmaz}
 \tilde{H}_C = \sum_{\alpha=1}^{r} J_\alpha \tau^x_\alpha + \sum_{\alpha=r+1}^{M} J_\alpha \prod_{\beta=1}^{r} (\tau^x_\beta)^{F_{\alpha,\beta}}.
\ee

\subsubsection{Finding $\tau^z$ operators}
Operators $\tau^z_\beta$ can be constructed using matrix $Z$ defined in Eq.~(\ref{Eq:Z}) in a way similar to how operators $\tau^x$ were constructed above. Specifically, we set
\begin{equation}\label{eq:tauz}
		\tau^z_\alpha \equiv \bigotimes_{l=1}^{N} (\sigma^x_l)^{Z_{\alpha, l}},  \quad \forall \alpha \in [1, r].
\end{equation}
The important difference is that vectors $z_\alpha$ contained in matrix $Z$ may contain a different number of non-zero entries. The commutation and anti-commutation properties of the $\{ \tau^z_\alpha, \tau^x_\beta\}$ operators follows directly from the orthogonality properties between $z_\alpha$ and $v_{\beta}$ vectors
\begin{equation*}
z_\alpha \cdot v_\beta =\delta_{\alpha \beta} \Rightarrow \left\{
\begin{split}
	\{\tau^z_\alpha,\tau^x_\alpha \} & = 0, \\ 
	[\tau^z_\alpha,\tau^x_{\beta}] & = 0 \hspace{0.2cm} \text{for} \; \alpha \neq \beta\in [1,r].
\end{split}
\right.
\end{equation*}

To find the dual operator of $H_X = \sum_i \sigma^x_i$ we have to invert Eq.~\eqref{eq:tauz} and find an expression for $\sigma^x$ operators via $\tau^z$.  This inversion procedure is straightforward for first $r$ spins that correspond to the invertible submatrix of $S_A$. Thus operators $\sigma^x_i$  for $i \in [1, r]$ read:
	\begin{equation}
	\label{eq:sigmax_first}
	\sigma^x_i \equiv \prod_{l=1}^{N} (\tau^z_l)^{(S_A)_{l ,i}},  \quad \forall i \in [1, r].
	\end{equation}
To obtain an expression for remaining $\sigma^x_{r+i}$ with $i \in [1, N-r]$ we use the knowledge of conserved charges from Eq.~\eqref{eq:basis_ker} and find that
\begin{equation}
	\label{eq:sigmax_last}
	\sigma^x_{r+i} \equiv O_i \prod_{l=1}^N (\tau^z_{l})^{(S_A)_{l, r+i}},
\end{equation}
where the particular choice of conserved charges is used as dictated by definition of  $\cal O$ matrix in Eq.~(\ref{eq:basis_ker}):
\begin{equation}\label{Eq:charge-cal-O}
O_l =\prod_i (\sigma^x_i )^{  \mathcal{O}_{l,i}}.
\end{equation}

\subsubsection{Dual Hamiltonian}	
Finally, joining Eq.~\eqref{eq:sigmaz}, ~\eqref{eq:sigmax_first} and ~\eqref{eq:sigmax_last}, we obtain the expression for the dual Hamiltonian
	\begin{multline}\label{eq:dualHf}	
		\tilde{H}_T(s) = -s\Big(\sum_{\alpha=1}^{r} J_\alpha \tau^x_\alpha + \sum_{\alpha=r+1}^{M} J_\alpha \prod_{\beta=1}^{r} (\tau^x_\beta)^{F_{\alpha,\beta}}\Big)  \\
		-(1-s)\Big(\sum_{i=1}^{r} \prod_{l=1}^{N} (\tau^z_l)^{(S_A)_{l ,i}} + \sum_{i=1}^{N-r}O_i  \prod_{l=1}^N (\tau^z_{l})^{(S_A)_{l, r+i}} \Big).
	\end{multline}

\subsection{Example}
Let us now illustrate the abstract procedure defined above using a specific example. We start from the matrix $\H$ corresponding to an instance of the 2-regular 3-XORSAT model with $N=6$ and $M=4$. The example considered here is a particular instance of the closure of the tree hypergraph Fig.~\ref{Fig:closedL}(a) with $g=1$, corresponding to the following $\H$ matrix:
	\begin{equation*}
	\H = \left(
	\begin{array}{cccccc}
		1 & 1 & 1 & 0 & 0 & 0 \\
		1 & 0 & 0 & 1 & 0 & 1 \\
		0 & 1 & 0 & 1 & 1 & 0 \\
		0 & 0 & 1 & 0 & 1 & 1 \\
	\end{array}
	\right).
	\end{equation*}
Similar to the main text we restrict to the case with all couplings $J_\alpha = 1$.

For this particular case, it is easy to check that the rank mod 2 of $\H$ is $r = 3$. To see this, for example, we could realize that the first row is the sum (mod 2) of all the other rows. We then pick a submatrix of $A$ containing all linearly independent rows as:
$$S_\H = (v_2, v_3, v_4 )^T = \left(
	\begin{array}{cccccc}
		1 & 0 & 0 & 1 & 0 & 1 \\
		0 & 1 & 0 & 1 & 1 & 0 \\
		0 & 0 & 1 & 0 & 1 & 1 \\
	\end{array}
	\right).
$$ 
Using Eq.~\eqref{Eq:taux-1} we then obtain:
	\begin{equation*}
		\tau^x_1 \equiv \sigma^z_1 \sigma^z_4 \sigma^z_6, \quad
		\tau^x_2  \equiv \sigma^z_2 \sigma^z_4 \sigma^z_5, \quad
		\tau^x_3 \equiv \sigma^z_3 \sigma^z_5 \sigma^z_6 .
	\end{equation*}

The $F$ matrix in this case corresponds to a row vector with all $r$ components being equal to one, $F = (1, 1, 1)$. Using Eq.~\eqref{Eq:taux-2} we obtain 
$$
\prod_{i = 1}^{3} \tau^x_i = \sigma^z_1\sigma^z_2\sigma^z_3.
$$
Hence, we can write the dual form of the classical Hamiltonian $H_C$ which reads
$$
\tilde{H}_X = \tau^x_1 + \tau^x_2 + \tau^x_2 + \tau^x_1\tau^x_2\tau^x_3.
$$
	
We now focus on defining the $\tau^z_\alpha$ operators. For this particular case, it is easy to check that 
$$Z=\left(
	\begin{array}{cccccc}
		1 & 0 & 0 & 0 & 0 & 0 \\
		0 & 1 & 0 & 0 & 0 & 0 \\
		0 & 0 & 1 & 0 & 0 & 0 \\
	\end{array}
	\right).
$$
Using Eq.~(\ref{eq:tauz}), we get
	\begin{equation*}
		\tau^z_\alpha \equiv \sigma^x_\alpha, \quad \forall \alpha = 1,2,3.
	\end{equation*}
From the above point, we can already read the expression for the $\sigma^x_i$ operators in terms of the $\tau^z_i$ operators for $i=1,2,3$. Furthermore, to find the expression of the remaining $\sigma^x_i$ operators ($i=4,5,6$) we need to find the conserved charges of the theory. 
	
Computing the kernel (mod 2) of $S_A$ we obtain:
	\begin{equation*}
		\mathcal{O}= \left(
\begin{array}{cccccc}
 1 & 1 & 0 & 1 & 0 & 0 \\
 0 & 1 & 1 & 0 & 1 & 0 \\
 1 & 0 & 1 & 0 & 0 & 1 \\
\end{array}
\right),
	\end{equation*}
which in the spin language from Eq.~(\ref{Eq:charge-cal-O}) corresponds to
	\begin{equation}
	\label{eq:ccharges_example}
		O_1 = \sigma^x_1 \sigma^x_2 \sigma^x_4, \quad
		O_2 = \sigma^x_2 \sigma^x_3 \sigma^x_5, \quad
		O_3 = \sigma^x_1 \sigma^x_3 \sigma^x_6.  
	\end{equation}
Thus, it only remains to find the set of clauses in which the spins $i=4,5,6$ participate. From that and using Eq.~\eqref{eq:ccharges_example}, we find the dual expressions for the remaining $\sigma^x $ operators:
	\begin{equation*}
		\sigma^x_4 \equiv O_4 \tau^z_1 \tau^z_2, \quad
		\sigma^x_5 \equiv O_5 \tau^z_2 \tau^z_3, \quad
		\sigma^x_6 \equiv O_6 \tau^z_1 \tau^z_3.		 
	\end{equation*}
	
The dual Hamiltonian follows directly from all the above results
	\begin{multline}
		H_T(s ) = -s \Big( \sum_{\alpha=1}^3 \tau^x_\alpha +  \tau^x_1 \tau^x_2 \tau^x_3\Big) \\
		-(1-s) \Big(\sum_{\alpha=1}^{3} \tau^z_\alpha + O_4 \tau^z_1\tau^z_2 +O_5 \tau^z_2\tau^z_3+O_6 \tau^z_1\tau^z_3\Big).
	\end{multline}
\section{Ising on the closure of the tree hypergraph} 
\label{app:ising_closed_lattice}

In this appendix, we provide details on the procedure that allows removing the non-local term $\tau^x_M$ in the dual Hamiltonian~(\ref{eq:DualHam-final}). The approach we present here is inspired by the one carried out in Ref.~\onlinecite{PRA_embedding_2016}. We engineer a Hamiltonian $\tilde{K}_T$ with an Abelian $Z_2$ symmetry, which is equivalent to $\tilde{H}_T$ in a given symmetry sector (which we denote as physical subspace). In addition, we request that in $\tilde{K}_T$ the non-local term becomes equivalent to a single spin operator. For this, we define the following projector
$P_0 =\sum_{\mathop{\bf{v}} \in \{0,1\}^{M-1}} |\mathop{\bf{v}_+} \rangle \langle \mathop{\bf{v}_+}|$, 
where 
$$
|\mathop{\bf{v}_+}\rangle = \frac{1}{\sqrt{2}} \big( |\mathop{\bf{v}},0\rangle  + |\mathop{\bf{v}},1\rangle \big).
$$
As a result, it is easy to check the following relations holds:
\begin{equation}
\label{eq:embedding}
\begin{split}
&P_0 \big(\tau^x_\alpha \otimes 1 \big)P_0 = X_\alpha, \; \alpha \in [1,M-1], \\
&P_0 \big(\prod_{\alpha=1}^{M-1}\tau^x_\alpha \otimes 1 \big) P_0 = X_M. \\
%&X_i \in \mathcal{H}_M \simeq \mathbb{C}^{\otimes M}_2, \; \forall i \in [1, M].
\end{split}
\end{equation}
Eq.~\eqref{eq:embedding} indicates that when restricted to the physical subspace the action of the $X_i$ operators for $i\in [1,M-1]$ is identical to that of the $\tau^x_i$ operators. On the other hand, the non-local term $\prod_{i = 1}^{M-1} \tau^x_i$ is now encoded in the $X_M$ operator associated with a new degree of freedom. We then have all the information needed to construct the Hamiltonian $\tilde{K}_T$.

We note that Eq.~\eqref{eq:embedding} directly implies that $\prod_{i=1}^M X_i = 1$, which completely specifies the physical subspace. Furthermore, the form that the remaining terms of $\tilde{H}_T$ take can be obtained from their (anti)commutation relations with the non-local term $\prod_{i = 1}^{M-1} \tau^x_i$. More specifically, we note that for operators $O_c$ commuting with the non-local operator the following holds
\be
P_0 \big([O_{c}, \prod_{i=1}^{M-1}\tau^x_i] \otimes 1 \big) P_0 = [P_0\big( O_{c} \otimes 1 \big)  P_0, X_M] = 0.
\ee
Hence, it implies that $P_0 \big( O_{c} \otimes 1 \big)P_0$  contains either the $X_M$ operator or acts as the identity on the spin $M$. 
However, using the definition of $P_0$ we see that only the identity on spin $M$ is permitted. In the same spirit, we see that for operators $O_{ac}$ anticommuting with the non-local term it holds that 
\be
P_0 \big( \{ O_{ac}, \prod_{i=1}^{M-1}\tau^x_i \} \otimes 1 \big) P_0 = \{ P_0 \big(O_{ac}\otimes 1 \big) P_0, X_M \} = 0, \\ 
\ee
which in turns implies that $P_0 \big( O_{ac}\otimes 1 \big) P_0$ has to contain either the $Z_M$ operator or the $Y_M$ operator. Using again the definition of $P_0$ we see that only the $Z_M$ operator is permitted. In this way, $\tilde{K}_T$ takes a form of transverse-field Ising model on the closed lattice, Fig.~\ref{Fig:closedL}:
\be
\tilde{K}_T(s) = -s \sum_{\alpha=1}^M J_\alpha X_\alpha - (1-s)\sum_{ \langle \alpha,\beta \rangle} Z_\alpha Z_\beta.
\ee
As a consistency check, we note that the subspace specified by the constraint $\prod_{\alpha=1}^{M} X_\alpha=1$ corresponds to the positive parity sector of $\tilde{K}_T(s)$ Hamiltonian with respect to $Z_2$ symmetry implemented by the operator $\prod_\alpha X_\alpha$. 

\end{document}